\documentclass{article}

\usepackage[a4paper, margin=0.6in]{geometry}
\usepackage{subcaption}
\usepackage[utf8]{inputenc}
\usepackage[dvipsnames,table,xcdraw]{xcolor}
\usepackage[affil-it]{authblk}
\usepackage[misc]{ifsym}
\usepackage{appendix,verbatim,listings,braket,enumitem, kantlipsum,graphicx,multicol,mathtools,csquotes,amsmath,mathtools,url,subfiles,qcircuit,amsthm,amssymb,fdsymbol,rotating,caption,hyperref,fontawesome,xcolor,arydshln,ulem,soul}

\def\titlefont{\color{Sepia}}

\let\OLDthebibliography\thebibliography
\renewcommand\thebibliography[1]{
  \OLDthebibliography{#1}
  \setlength{\parskip}{0pt}
 \setlength{\itemsep}{0pt plus 0.3ex}
}

\title{\titlefont \textbf{\LARGE A resource-efficient variational quantum algorithm for \\mRNA codon optimization}}

\author[1, \Letter]{Hongfeng Zhang}
\author[2]{Aritra Sarkar}
\author[1]{Koen Bertels}

\affil[1]{Department of Electronics and Information Systems, University of Ghent, Belgium}
\affil[2]{Department of Quantum \& Computer Engineering, Delft University of Technology, The Netherlands}

\affil[ \Letter ]{Hongfeng.Zhang@UGent.be}

\date{}

\begin{document}

\maketitle

\begin{abstract}

Optimizing the mRNA codon has an essential impact on gene expression for a specific target protein. 
It is an NP-hard problem; thus, exact solutions to such optimization problems become computationally intractable for realistic problem sizes on both classical and quantum computers.
However, approximate solutions via heuristics can substantially impact the application they enable. 
Quantum approximate optimization is an alternative computation paradigm promising for tackling such problems.
Recently, there has been some research in quantum algorithms for bioinformatics, specifically for mRNA codon optimization.
This research presents a denser way to encode codons for implementing mRNA codon optimization via the variational quantum eigensolver algorithms on a gate-based quantum computer.
This reduces the qubit requirement by half compared to the existing quantum approach, thus allowing longer sequences to be executed on existing quantum processors.
The performance of the proposed algorithm is evaluated by comparing its results to exact solutions, showing well-matching results. 
\end{abstract}


\section{Introduction} \label{sec:intro}

Bioinformatics plays a pivotal role in various domains of biology, including personal medicine, drug discovery, genome assembly, genetically modified crops, and vaccine development, all of which have a profound impact on individuals' lives. 
This interdisciplinary field seamlessly merges biology and computer science, serving as a catalyst for addressing pertinent challenges in biology and bio-medicine. 
An extensive volume of biological data is rapidly generated because of the advent of high-throughput experiments. 
Bioinformatics takes on the formidable task of analyzing and interpreting these vast datasets, thereby furnishing solutions to intricate biological quandaries. 
Consequently, there is an imperative need for efficient data processing to bolster production and advancements within this field. 
It's important to acknowledge the inherent complexity of specific bioinformatics problems. 
For instance, challenges like protein folding, mRNA codon optimization, and mRNA folding are characterized by their combinatorial nature and NP-hard problems~\cite{berger1998protein}\cite{csen2020codon}\cite{fox2021mrna}. 
As the volume of biological data continues to surge, it exacerbates the computational complexity of these issues. Nevertheless, bioinformatics remains a critical driving force in advancing biology and bio-medicine, continually striving to tackle these challenges and unlock new possibilities for the benefit of humanity.

In bioinformatics, computational problems often present significant challenges due to the massive amount of data and the substantial computational resources. Searching for effective solutions to these problems is important for the development of bioinformatics. Quantum computing emerges as a promising candidate due to its distinctive properties, including superposition, entanglement, and interference between computational paths. These properties enable quantum computers to process vast amounts of data in parallel, potentially overcoming the limitations of classical computing for bioinformatics tasks. Therefore, exploring quantum algorithms for bioinformatics holds significant promise for accelerating the advancements in this field. 

The article is organized as follows.
Section~\ref{sec:related} provides a brief introduction, highlighting the profound impact of quantum computing on the field of bioinformatics. 
We also introduce our innovative approach, emphasizing its resource-efficient nature, as the solution to mRNA codon optimization. 
The associated terminologies of genomics, such as DNA, RNA, and synonymous codons, are introduced. 
In section~\ref{sec:method}, we meticulously delineate our methodology. 
We offer a detailed account of the innovative approach, providing insights into its core mechanisms. 
Section~\ref{sec:result} is dedicated to the presentation of our research results. 
We showcase the efficiency and effectiveness of our approach in solving the mRNA codon optimization challenge. 
The comparison of the required number of qubits on the existing approach \cite{fox2021mrna} and our one. mRNA codon optimization for protein P0DTC2 was executed on the Qiskit platform. 
We conclude the article in section~\ref{sec:conclusion}, encapsulating the essence of our study. 
We summarize our key findings and underscore the importance of implementing quantum computing algorithms in the realm of mRNA codon optimization.

\section{Related research} \label{sec:related}

In the realms of DNA or RNA vaccine development, bioinformatics leverages the analysis of extensive pathogen sequences to identify target antigenic proteins for mRNA sequence design. 
This expedites the vaccine design process, facilitating the creation of safe and effective vaccines tailored to combat specific viruses. 
The COVID-19 pandemic, which spread rapidly from 2019 to 2022, had a profound global impact, affecting a vast number of people. 
This surge strained healthcare systems, leading to a scarcity of health resources and thereby preventing many individuals from receiving timely treatment. 
This had wide-ranging consequences on global health and everyday life. 
Additionally, the pandemic triggered a surge in unemployment, trade disruptions, and a downturn in the global economy \cite{pak2020economic}.
This underscored the urgent need to conduct research into vaccine development, particularly focusing on highly infectious viruses, to safeguard human health.  

In the face of global health threats, the rapid development of effective vaccines is crucial to protect public health and mitigate the impact of such viruses. 
One of the promising candidates in this endeavor is mRNA vaccines. 
They offer advantages over traditional vaccines, such as safety, rapid development, high protein expression efficiency, and reduced production time \cite{pardi2018mrna}.
In what follows, we review the associated terminologies in genomics required for understanding mRNA vaccines. 

\begin{table}[bht]
    \centering
    \begin{tabular}{|c|c|c|c|}
         \hline
    
    \textbf{Amino acid} & \textbf{mRNA codons} & \textbf{Amino acid} & \textbf{mRNA codons}\\ 
    \hline
    Ala~(A)  & \verb|GCU, GCC, GCA, GCG|  & IIe~(I)  & \verb|AUU, AUC, AUA| \\ 
    \hline
    Arg~(R)  & \verb|CGU, CGC, CGA, CGG, AGA, AGG| & Leu~(L) & \verb|CUU, CUC, CUA, CUG, UUA, UUG| \\ 
    \hline
    Asn(~N)  & \verb|AAU, AAC| & Lys~(K)  & \verb|AAA, AAG| \\
    \hline
    Asp~(D)  & \verb|GAU, GAC| & Met~(M)  & \verb|AUG| \\
    \hline
    Asn or Asp~(B)  & \verb|AAU, AAC, GAU, GAC| & Phe~(F)  & \verb|UUU, UUC| \\
    \hline
    Cys~(C)  & \verb|UGU, UGC| & Pro~(P)  & \verb|CCU, CCC, CCA, CCG| \\
    \hline
    Gln~(Q)  & \verb|CAA, CAG| & Ser~(S)  & \verb|UCU, UCC, UCA, UCG, AGU, AGC|\\
    \hline
    Glu~(E)   & \verb|GAA, GAG| & Thr~(T)  & \verb|ACU, ACC, ACA, ACG| \\
    \hline
    Gln or Glu~(Z)  & \verb|CAA, CAG, GAA, GAG| & Trp~(W)  & \verb|UGG| \\
    \hline
    Gly~(G)  & \verb|GGU, GGC, GGA, GGG| & Tyr~(Y)  & \verb|UAU, UAC| \\
    \hline
    His~(H)  & \verb|CAU, CAC| & Val~(V)  & \verb|GUU, GUC, GUA, GUG| \\
    \hline
    Start & \verb|AUG, CUG, UUG| & Stop & \verb|UAA, UGA, UAG| \\
    \hline
    \end{tabular}
    \caption{mRNA codon table from \cite{enwiki:codon_table}} 
    \label{tab:my_label}
\end{table}

\begin{figure}[htb]
    \centering
    \includegraphics[width=0.65\textwidth]{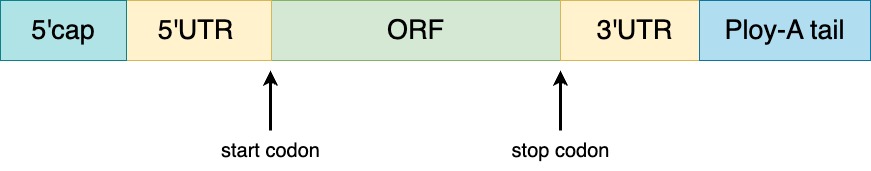}
    \caption{The structure of mRNA sequence}
    \label{fig:structure_mrna}
\end{figure}

\begin{itemize}
    \item \textbf{mRNA -} Messenger RNA(mRNA), on the other hand, serves as the messenger between DNA and proteins, carrying the genetic code and orchestrating the protein-building process. mRNA consists of a sequence of nucleotides, including adenine (A), cytosine (C), guanine (G), and uracil (U). Transcription translates DNA into mRNA, which subsequently deciphers this code to construct proteins. The reverse process, translation, is pivotal in the assembly of amino acids into a polypeptide chain.
    \item \textbf{Codon and synonymous codon -} A codon is a sequence of three consecutive nucleotides in RNA or DNA, and it encodes a specific amino acid. The fact that there are only 4 different nucleotides (A, C, G, T in DNA and A, C, G, U in RNA) means that 64 possible codons ($4^3=64$) can be formed. Each of these codons can represent a specific amino acid in the genetic code. It's evident that there are more codons than unique amino acids as there are only 20 types of amino acids. As a result, some codons encode the same amino acid in Table \ref{tab:my_label} from \cite{enwiki:codon_table}. These codons that can be mapped to the same amino acid are known as \textbf{synonymous codons}. For example, the amino acid Serine~(S) is encoded by one of six synonymous codons: UCU, UCC, UCA, AGU, AGC, or UCG. Figure \ref{fig:mRNA_polypeptide} represents the relationship between a mRNA sequence and a polypeptide chain where the part of an amino acid chain, YSFLV, is decided by an mRNA sequence, UAU-UCU-UUU-CUU-GUU.

    \begin{figure}[htb]
        \centering
        \includegraphics[width=0.7\textwidth]{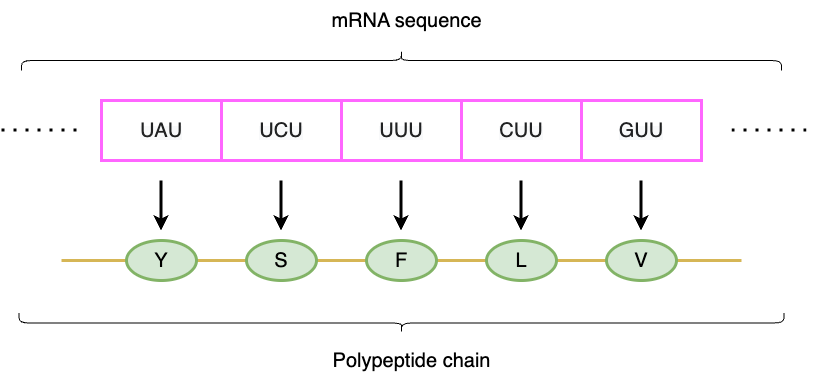}
        \caption{the transcription between a mRNA sequence and a polypeptide chain}
        \label{fig:mRNA_polypeptide}
    \end{figure}

    \item \textbf{mRNA vaccines -} mRNA vaccines have emerged as a revolutionary approach to immunization, utilizing a fragment of messenger RNA (mRNA) to trigger the production of viral proteins (antigens) within the human body. These antigens, produced by human cells in response to the injected mRNA, stimulate the immune system to generate protective antibodies. This cutting-edge technique not only offers safety advantages by not using weakened or dead viruses but also exhibits rapid development, high protein expression efficiency, and shortened production timelines \cite{pardi2018mrna}\cite{schlake2012developing}\cite{pardi2020recent}. 
    \item \textbf{Structure of the mRNA sequence -} As shown in Figure \ref{fig:structure_mrna}, the structure of mRNA sequences is composed of the following components: \textbf{5' cap, 5' Untranslated  Region(5' UTR), Open Reading Frame(ORF), 3' Untranslated Region (UTR)}, and \textbf{Ploy-A tail} \cite{meijer2002control}. While the 5' untranslated region (5'UTR) and the 3' untranslated region (3'UTR) are essential in regulating protein expression, they do not directly determine the specific amino acids incorporated into the protein. It's worth noting that this paper does not discuss the optimization of mRNA codons through the 5'UTR and 3'UTR regions. The ORF is a part of the DNA, RNA, or mRNA sequence, which begins with a start codon (usually AUG, CUG, UUG in mRNA) and ends with a stop codon(UAA, UAG, and UGA in mRNA). The ORF in mRNA serves as the blueprint for protein synthesis, dictating the specific sequence of amino acids to be incorporated with the resulting antigen protein. In addition, the ORF is crucial for the development of mRNA vaccines, as it also plays a pivotal role in protein expression and stability \cite{tokuoka2008codon}\cite{mauro2018codon}.
    
\end{itemize}
This paper focuses on codon optimization of ORF, a method aimed at enhancing the efficiency and performance of mRNA vaccines. Codon optimization includes identifying the most optimal combination of synonymous codons for a given protein sequence. 

However, several challenges persist in optimizing the sequence of ORF for mRNA vaccines. As highlighted in \cite{quax2015codon}, the issue of designing mRNA sequences is inherently a combinatorial optimization problem. 
The solution space grows exponentially when the length of the sequence increases. 
Despite the development of some classical computing methods(heuristics, metaheuristics, and exact algorithms), no existing tools offer high-performance capabilities in mRNA sequence design. 
Given the urgency of rapidly producing mRNA vaccines upon the identification of a virus, there is a compelling need to devise a new computing model or algorithm that can efficiently address this challenge. 
Such an innovation would not only reduce the time spent on mRNA sequence design but also expedite the administration of mRNA vaccines to safeguard public health. mRNA codon optimization can be approached in different ways, broadly categorized into two types, namely the classical and quantum approaches.

\begin{itemize}
    \item \textbf{Classical approach to mRNA vaccine development} 
    
    Traditional methods for solving this problem on classical computers can be computationally intensive, especially as the length of the target sequence increases. The complexity arises from the vast search space of possible codon combinations and the need to balance factors like codon usage bias and secondary mRNA structure. Deep learning, particularly neural networks, has shown promise in addressing complex bioinformatics problems, including mRNA codon optimization \cite{fu2020codon}\cite{gong2023integrated}\cite{zhang2023algorithm}. Neural networks can automatically learn meaningful representations of biological data. In the context of mRNA codon optimization, this means that the model can capture the essential features of codon sequences that lead to optimal protein expression. Pre-trained models on related tasks or datasets can be fine-tuned for specific mRNA codon optimization problems. This can leverage knowledge gained from one domain to enhance performance in another. While deep learning holds promise, it's essential to note that the success of these models depends on the availability of high-quality training data and careful design of the neural network architecture. Additionally, interoperability and understanding the biological implications of the model's predictions are crucial in the context of bioinformatics applications.
    
    \item \textbf{Quantum approach to mRNA vaccine development} 
    
    Quantum computers have the potential to handle combinatorial optimization tasks because of their use of superposition and entanglement to parallel process quantum states. Quantum computing is able to achieve significant speedup compared to classical computing for addressing combinatorial optimization, as it can explore solutions in superposition to find the optimal result faster. It can adeptly tackle combinatorial problems within vast solution spaces, making it a promising candidate for fields like cryptography, artificial intelligence, and genomics, where significant computational power is required for expedited processes. Quantum computing has already been attended in bioinformatics, where it's being applied to genome sequence reconstruction algorithms \cite{sarkar2021qibam,sarkar2021quaser}, genome sequence analysis \cite{sarkar2021estimating}, protein folding \cite{robert2021resource}, and mRNA codon optimization \cite{fox2021mrna}. 
    
    \begin{itemize}
        \item \textbf{Quantum Annealing (QA)} 
        
        Quantum annealers, like those developed by D-Wave Systems, hold significant promise for solving combinatorial optimization problems \cite{kadowaki1998quantum}\cite{santoro2006optimization}\cite{das2008colloquium}. Quantum annealing is utilized for practical applications, including in the field of mRNA codon optimization. \cite{fox2021mrna} The general process of applying quantum annealer to combinatorial optimization problems, such as mRNA codon optimization involves several steps. One is problem definition, specifying the optimization goal and relevant constraints. The second is to transform the problem into an Ising model representation. Then, the quantum annealing devices are used to find the lowest energy state. The final step is to decode the result to obtain the mRNA codon sequence, reflecting an optimization solution for protein expression. 
        
        \item \textbf{Variational Quantum Eigensolver (VQE)} 
        
        VQE is a quantum algorithm designed to find the minimum eigenvalue for a given Hamiltonian, making it well-suited for solving certain types of combinatorial problems \cite{peruzzo2014variational}. Firstly, a combinatorial problem is defined using a Hamiltonian. Then, a circuit ansatz with variational parameters is optimized to the approximate ground state of the Hamiltonian corresponding to the minimum expectation value. These quantum circuits are executed on a quantum computer to obtain the expectation value of the Hamiltonian. The process is iterated until convergence or a satisfactory solution quality is reached. 
        
        \item \textbf{Quantum Approximate Optimization Algorithm (QAOA)} 
        
        QAOA is a variational quantum algorithm, belonging to a broader class of VQE. It is tailored to find approximate optimization problems' solutions \cite{farhi2014quantum}.  The algorithm involves mapping a specific problem to the Ising model. In the aspect of quantum circuit construction, quantum state preparation encodes potential solutions, and a quantum circuit with alternating layers of unitary operations is built. Parameters are adjusted using classical optimization techniques to minimize the expectation value of the Hamiltonian. Iterations continue until convergence is reached.  
    \end{itemize}   
\end{itemize}

We have developed an innovative quantum method that significantly reduced quantum computing resources with a dense way of encoding compared to the existing quantum approach \cite{fox2021mrna}. However, our approach adds complexities to the Hamiltonian due to the multiple local terms. Quadratic unconstrained binary optimization (QUBO) and QAOA approach mandates a Hamiltonian composed of only two local terms and it needs to add extra qubits for the reduction of additional terms. But this reduction also results in an increase in the required number of qubits. To address these complexities, we have employed the VQE algorithm, as it offers a unique advantage: it does not demand the addition of qubits. VQE enables us to seek the ground energy of the objective function efficiently, following which we employ classical computing techniques to derive the optimal mRNA sequence based on the ground state.
    
\textbf{Qiskit platform --} Our implementation of codon optimization is conducted using the Qiskit platform which is the gate circuit quantum simulator as a central tool for codon optimization. Nevertheless, current quantum computing resources are constrained by the number of qubits. For example, optimizing the full-length protein sequences would demand thousands of qubits, surpassing the capabilities of existing quantum hardware. We've adopted a pragmatic approach to overcome this limitation by dividing the protein sequence into smaller, more manageable fragments with shorter lengths. This strategic division enables us to operate with fewer qubits for each fragment, ensuring a more feasible and efficient optimization process. By presenting this innovative approach to mRNA codon optimization, we aim to contribute to the evolving field of quantum computing in bioinformatics, providing new perspectives and solutions to complex problems in the realm of genetic coding and sequence optimization. 
The Qiskit platform allows to execution of quantum circuits in two ways. The first one connects to a native QPU backend like superconducting qubits, taking the results from the latest quantum processors into account. The second one is using a quantum circuit simulator where the qubits are assumed to be perfect qubits. That intrinsically implies no decoherence of the qubits and that the quantum gates have no error. Error models can be applied on the simulator for realistic simulation, however, often become a barrier to quantum logic and algorithm development for future quantum applications. \textbf{For this project, we employ noiseless quantum circuit simulation for the proof-of-concept development and demonstration of our algorithm, based on the Perfect Intermediate Scale Quantum~(PISQ)~\cite{bertels2021quantum} approach.}

\section{Method} \label{sec:method}

Optimizing the mRNA sequence involves the meticulous selection of codons across the entire amino acid chain. Remarkably, different codon choices within an mRNA sequence can significantly impact gene expression, protein folding, and mRNA stability. The core of this endeavor lies in mRNA codon optimization, a complex task that requires the arrangement of synonymous codons in an optimal combination for a specific protein sequence.

This undertaking unfolds as a combinatorial optimization problem, with complexity that escalates exponentially as the length of polypeptide chains increases. Each position in a polypeptide chain has at most $6$ choices of synonymous codons.  A polypeptide chain that consists of $N$ position (amino acids) may entail up to $6^N$ possible codon combinations. This expansive solution space becomes particularly formidable when $N$ equals $200$, resulting in a solution space numbering approximately $6^{200} \approx 3.1 * 10^{151}$, which is astronomically large.

The quest for optimizing mRNA sequence within this vast solution space, for a particular bioinformatics application, poses a formidable challenge that necessitates significant computational resources. To tackle this challenge, two approaches have been explored: the development of new algorithms, encompassing heuristic methods, machine learning, and Monte Carlo techniques, and the utilization of high-performance computing technologies, including parallel computing, cloud computing, and quantum computing.

Quantum computers, in particular, hold promise due to their distinctive quantum mechanical properties such as superposition, entanglement, and interference. These properties empower quantum computers to efficiently handle extensive quantum data sets, finding the optimal (lowest or highest) value in a vast space solution. Notably, Fox et al. \cite{fox2021mrna} introduced a codon optimization algorithm implemented on quantum annealers and gate-based circuit quantum computers. 
This groundbreaking work offers a practical approach to mRNA sequence design, unlocking new possibilities in the realm of bioinformatics and genetic research. The research by these authors represents a crucial milestone in harnessing the capabilities of quantum computing for this innovative application. 

Our quantum algorithm for mRNA codon optimization focuses on enhancing efficiency with resource optimization. It enables efficient optimization for longer mRNA sequences than the reference work. Our quantum method involves three steps, encoding synonymous codons, constructing Hamiltonian, and executing quantum circuits. The approach is pivotal in mRNA vaccine development, leveraging quantum computing advantages to address the combinatorial problem efficiently. It presents a groundbreaking solution for achieving optimal mRNA design.

\subsection{Synonymous codons encoding}
The first step is to encode synonymous codons using a string of qubits for the target protein chain, ensuring each synonymous codon has a unique representation. This is a crucial step for mRNA codon optimization because it directly relates to the required number of qubits. An efficient encoding way can lead to a reduction in the required number of qubits, which is essential for ensuring the efficiency of quantum computing. One-hot encoding is used by the existing method \cite{fox2021mrna}, and the dense encoding is adapted by our approach.

\begin{itemize}
    \item \textbf{One-hot encoding:} One-hot encoding utilizes a binary string where a single '1' is placed at the position corresponding to the represented codon. The rest of the positions contain '0's, ensuring that only one position is active. The most common encoding way of quantum computing is one-hot encoding, where a unique string of bits is used to represent a synonymous codon of an amino acid. In this approach, a string of qubits represents one amino acid, with only one position set to 1 to indicate the chosen codon. When a specific bit pattern appears in the final optimal result, it signifies the selected codon for the mRNA sequence. For example, consider the amino acid Leucine (Leu), which has six synonymous codons: UUA, UUG, CUA, CUC, CUG, and CUU so these codons are mapped to binary representations, such as 000001, 000010, 000100, 001000, 010000, and 10000, respectively in one-hot encoding. The resulting qubit string is represented as $q_0 q_1 q_2 q_3 q_4 q_5 q_6$, where each $q_i$ can take on values of 0 or 1, but only one of the qubits is set to 1. 
    \item \textbf{Dense encoding:} In contrast to one-hot encoding, dense encoding uses binary bit strings representing synonymous codons. For instance, the six synonymous codons of Leucine (Leu) are encoded as 000, 001, 010, 011, 100, and 101, respectively. In this encoding scheme, the qubits for each codon are represented as $q_0 q_1 q_2$, where each $q_i$ can be either 0 or 1. The critical advantage of dense binary encoding is that it requires fewer qubits compared to one-hot encoding. However, it is crucial to ensure that the decoding process can correctly map these binary patterns to the corresponding codons. The example given in Table \ref{tab:example_A_L_encoding} outlines the encoding for amino acids A (Alanine) and L (Leucine) using one-hot and dense binary encoding. The table provides a clear reference for how each synonymous codon represents the one-hot and the binary pattern. 
\end{itemize}

Two examples for encoding amino acids A and L are given in  Table \ref{tab:example_A_L_encoding} using either the one-hot or the dense way. This encoding allows straightforward identification of the selected codon based on the binary pattern. However, it introduces redundant quantum states (explained in the Appendix), which in turn increases the complexity of the quantum system. 
\begin{table}
    \centering
    \begin{tabular}{|c|c|c|c|}
    \hline
        amino acid & codon & one-hot encoding qubits & dense encoding qubits \\
        \hline
        Ala~(A) & \verb|GCA| & 1000 & 00\\
          & \verb|GCC| & 0100 & 01\\
          & \verb|GCG| & 0010 & 10\\
          & \verb|GCU| & 0001 & 11\\
        \hline
        Leu~(L) & \verb|CUA| & 100000 & 000\\
          & \verb|CUC| & 010000 & 001\\
          & \verb|CUG| & 001000 & 010\\
          & \verb|CUU| & 000100 & 011\\
          & \verb|UUA| & 000010 & 100\\
          & \verb|UUG| & 000001 & 101\\
        \hline
    \end{tabular}
    \caption{Example for encoding amino acids, Ala~(A) and Leu~(L), into qubits with one-hot way and dense way}
    \label{tab:example_A_L_encoding}
\end{table}
The two encoding methods yield distinct qubit requirements for the same protein chain. As shown in Equation \ref{one hot}, one-hot encoding requires a number of qubits equivalent to the count of synonymous codons for each amino acid. In contrast, as denoted in Equation \ref{dense}, the dense encoding method typically calls for approximately half the number of qubits compared to one-hot encoding. This discrepancy in qubit utilization demonstrates the efficiency gains achievable with the dense encoding approach.\\
\begin{equation}
    \label{one hot}
    C_{one-hot} = \sum_{i=0}^{N-1} C_{i}
\end{equation}
\begin{equation}
    \label{dense}
    C_{dense} = \sum_{i=0}^{N-1}\lceil \log C_{i} \rceil
\end{equation}
where $C_i$ represents the number of synonymous codons at ith amino acid on a protein chain, and N represents the number of amino acids in a protein sequence. 

\subsection{Hamiltonian construction }

This involves the construction of a Hamiltonian, a key component for optimizing the codons of any protein sequence. The Hamiltonian comprises various factors influencing the mRNA-codon optimization, including codon usage bias, target GC concentration, and sequentially repeated nucleotides. Additionally, it encompasses constraint terms related to redundant encoding, which cannot be mapped to an actual mRNA sequence (explained in the Appendix). The ultimate objective is to identify the ground state of the Hamiltonian, which enables the decoding process to obtain the optimal mRNA sequence. The Hamiltonian of mRNA codon optimization is the equation \ref{Hamiltonian_method} given in \cite{fox2021mrna}.
\begin{equation}
\label{Hamiltonian_method}
    H(q) = H_f(q) + H_{gc}(q) + H_r(q) + H_p(q)
\end{equation}
where $H_f$ represents codon usage bias, $H_{gc}$ represents target gc concentration, $H_r$ represents sequentially repeated nucleotides, and $H_p$ represents the constraint items. All of the explanations about $H_f, H_{gc}, H_r, \text{ and } H_p$ and how to construct the Hamiltonian are given in the Appendix.

\subsection{Quantum algorithm execution}
In the case of one-hot encoding, the Hamiltonian exhibits only one and two local interaction terms, making it readily translatable into a Binary Quadratic Model (BQM) suitable for QAOA and QUBO approaches. However, the Hamiltonian associated with the dense encoding method (as shown in Equation \ref{Hamiltonian}) incorporates more than two interaction terms. To adapt it for QUBO or QAOA, additional qubits must be introduced to reduce the multiple interaction terms, leading to an increase in the required number of qubits. VQE is commonly used to search for the ground energy, without necessarily focusing on the corresponding state. Nevertheless, in this context, the objective is to identify the state with the lowest ground value of the Hamiltonian. This specific state, featuring a locally optimal ground value, or the best ground value, can then be translated into an optimal mRNA sequence.

The VQE's output comprises the minimum energy value and the angles associated with the ansatz. Subsequently, the states are generated based on the ansatz angles, and the state with the lowest energy is the target for codon optimization. Listing all possible states is unnecessary, as many of them are irrelevant and do not contribute to minimizing the Hamiltonian. It is more efficient to set a probability threshold and list states with higher probabilities to identify the optimal state. Ultimately, this optimal state can be transformed into the desired mRNA sequence, representing the desired outcome of the codon optimization process.

The SamplingVQE on the Qiskit platform \cite{barkoutsos2020improving} was used to implement the mRNA codon optimization for the protein fragment with a length of 8 while utilizing between 8 and 21 qubits. 
Given the challenges in accessing real quantum computers with a sufficient number of qubits, simulators in Qiskit allow scientists to explore optimization problems in quantum computing. In addition, Python3 served to generate Hamiltonian and as an accessible programming language on Qiskit. In the final step, the outcome of Qiskit was translated to mRNA sequence through Python3.

\section{Results} \label{sec:result}

The protein P0DTC2 was an example of optimizing mRNA codon sequences. The P0DTC2 protein from the spike protein of 2019-nCoV, sourced from UniProt and consisting of 1273 amino acids, would require 2234 and 4417 qubits for dense and one-hot encoding respectively according to equations \ref{one hot} and \ref{dense}. However, the available number of qubits on current quantum computers is limited, with no quantum computer possessing more than 1000 qubits. It is unrealistic to achieve mRNA codon optimization for the full-length protein P0DTC2 on a quantum computer because of the large number of qubits required. To overcome the limitation, the protein sequence is divided into shorter fragments, and each fragment is optimized separately by VQE. This approach requires fewer qubits for each fragment, making the optimization more manageable than optimizing the whole protein sequence at once. We use paper~\cite{fox2021mrna} as our reference paper.

\begin{figure}[th]
    \centering
    \begin{subfigure}[b]{0.5\textwidth}
        \includegraphics[width=\textwidth]{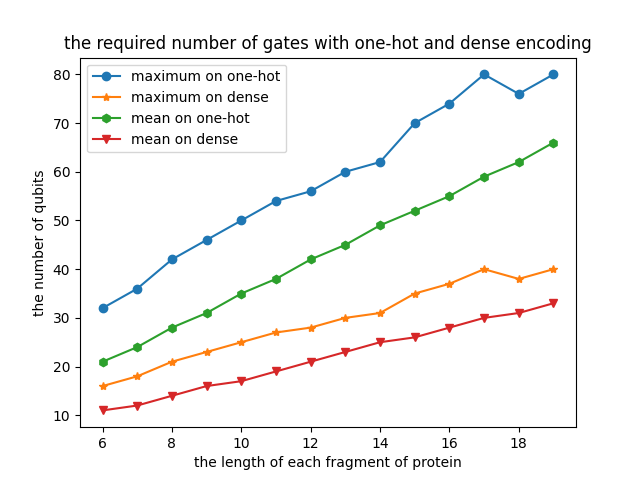}
        \caption{The number of qubits that protein P0DTC2 requires with one-hot and dense encoding}
        \label{fig:qubits_P0DTC2}
    \end{subfigure}
    \hfill
    \begin{subfigure}[b]{0.45\textwidth}
        \includegraphics[width=\textwidth]{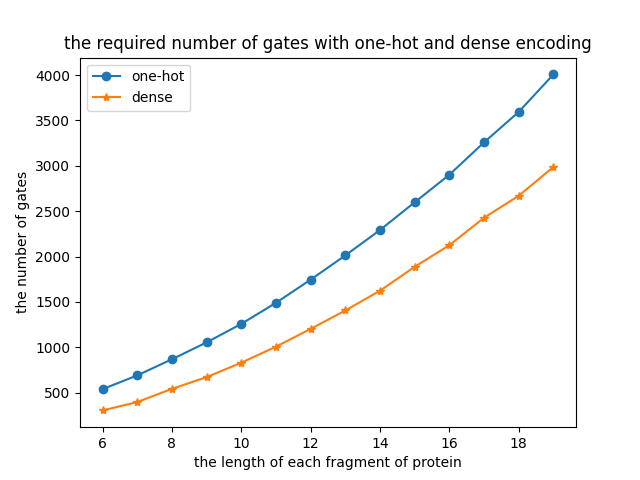}
        \caption{The required number of gates that protein P0DTC2 requires with one hot and dense encoding}
        \label{fig:gates_P0DTC2}
    \end{subfigure}
    
    \caption{The number of required qubits on protein P0DTC2}
    \label{fig:qubits_protein}
\end{figure}

    


In Figure \ref{fig:qubits_P0DTC2}, we extended this comparison to assess the maximum number of qubits required for encoding the P0DTC2 protein sequence using both one-hot and dense encoding across fragments with lengths ranging from 6 to 20 according to equations \ref{dense} and \ref{one hot}. For each fragment with a length of 6, the maximum number of qubits required for dense encoding is 16, whereas one-hot encoding requires 32. On average, dense encoding uses 11 qubits while one-hot needs 21. For longer fragments with a length of 19, the average qubit count for dense encoding is 33 compared to 66 with one-hot encoding. The maximum number of qubits for dense is 40 while one-hot requires 80 qubits. These results underscore the substantial reduction in the number of qubits needed when using dense encoding, which is significantly less than half of the requirement for one-hot encoding. This indicates that dense encoding is a resource-efficient approach for mRNA codon optimization in quantum computing compared to the current method \cite{fox2021mrna}. 

\begin{figure}[htb]
    \centering
    \includegraphics[width=0.5\textwidth]{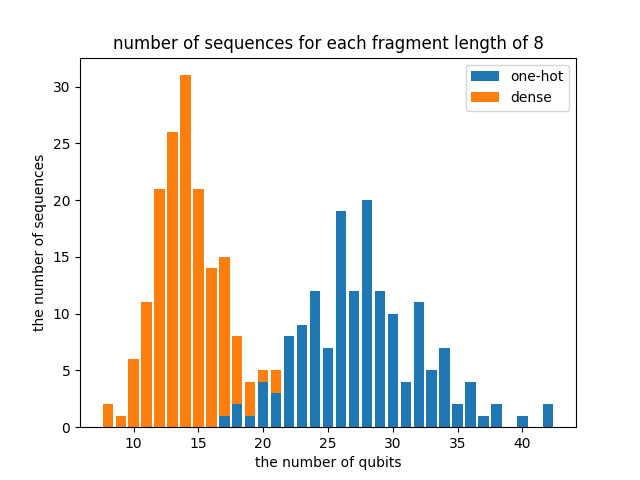}
    \caption{The required number of qubits distribution between one hot and dense encoding on protein P0DTC2}
    \label{fig:compare_qubit_8_P0DTC2}
\end{figure}

Figure \ref{fig:gates_P0DTC2} shows that the average number of quantum gates required per fragment for encoding on the P0DTC2 protein varies significantly depending on the encoding method - one-hot and dense. For short fragments with a length of 6, one-hot encoding requires 537 gates, whereas dense encoding only requires 303 gates. For fragments with a length of 19, 4000 gates are required by one-hot encoding while dense encoding requires approximately 3000 gates. It indicates that the dense method uses fewer gates than one-hot encoding.

In the paper, we divided the protein into 159 fragments, each with a length of 8 amino acids (excluding the first amino acid M, which has only one codon and does not need optimization). Consequently, the number of required qubits for each fragment falls within the range of 8 to 21, as determined by equation \ref{dense}. Each fragment is used to construct a Hamiltonian via equation \ref{Hamiltonian}, and then Sampling VQE on Qiskit is employed to identify the optimal codon sequence for each fragment. For each fragment with a length of 8, we conducted a comparison of the required number of qubits between one-hot and dense encoding for the protein P0DTC2. In Figure \ref{fig:compare_qubit_8_P0DTC2}, a comparison of the required number of qubits with one hot and dense encoding for every fragment with a fixed length of 8 on P0DTC2 can be computed by Python3 according to equations \ref{dense} and \ref{one hot}. The one-hot encoding requires a minimum of 16 qubits and goes up to 42 qubits for maximum representation. In contrast, dense encoding requires a minimum of 8 qubits and extends up to 21 qubits for maximum representation. Most fragments in one-hot and dense encoding utilize 14 and 28 qubits respectively. 
\begin{figure}
    \centering
    \includegraphics[width=0.5\textwidth]{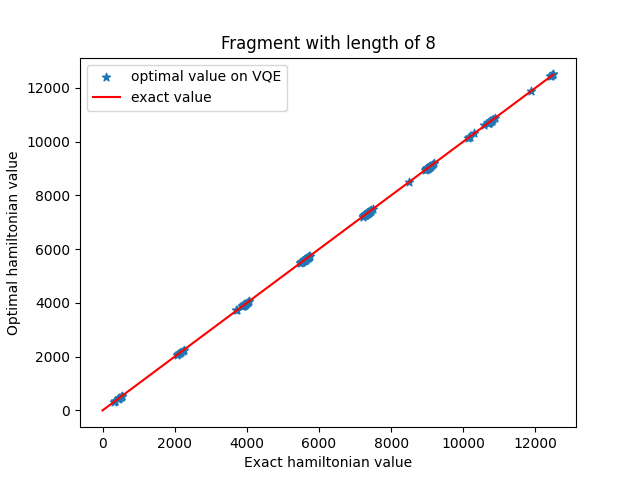}
    \caption{Exact and optimal value of protein P0DTC2 on Qiskit}
    \label{fig:P0DTC2_8}
\end{figure}

The main results of our method on Qiskit and the exact values are visualized in Figure \ref{fig:P0DTC2_8}. In this figure, the red line represents the exact value for each fragment, while each data point is the value generated by our approach in quantum computing. A noteworthy observation is that the majority of data points align closely with the line $y=x$. This alignment indicates that the sequences produced by our proposed algorithm are close to the best sequences, illustrating the accuracy of our approach.

In conclusion, this significant reduction in the required number of qubits and gates with dense encoding suggests potential advantages over one-hot encoding from Figure \ref{fig:qubits_P0DTC2} and \ref{fig:gates_P0DTC2}. In Figure \ref{fig:P0DTC2_8}, the quantum approach presented in this paper is designed to identify the optimal mRNA sequence by ensuring the Hamiltonian value matches the exact value that can be calculated by Python3. These results show that the quantum method has the capability to optimize mRNA codons more effectively than the current one \cite{fox2021mrna}.

\section{Conclusion} \label{sec:conclusion}

Our paper presents a compelling method to advance mRNA codon optimization through the innovative application of quantum computing - a critical frontier in mRNA vaccine development. The core motivation to enhance the efficiency and stability of mRNA vaccines is well-found, aligning with the contemporary progress in biotechnology and medicine. mRNA codon optimization is an NP problem and classical computers have limitations in tackling NP-hard problems. However, the unique quantum properties, such as parallelism and speedup, offer a promising avenue for unraveling intricacies associated with the pivotal task. The current constraints in quantum computing, including the limited number of realistic qubits and coherence time, reflect a pragmatic perspective. The focus on minimizing the required number of qubits constitutes a strategic and pivotal aspect, directly influencing the feasibility of the proposed quantum method. Our novel quantum method has abilities to handle longer fragments of protein chains on quantum computers compared to classical counterparts.

However, our paper acknowledges a limitation in selecting optimization factors, specifically codon usage bias, GC content, and sequentially repeated nucleotides. The suggestion that other potential factors influencing mRNA codon optimization might exist is insightful. Further investigation could delve into additional constraints, enriching the realism and comprehensiveness of mRNA codon optimization. Collaborating with experts in biology or related fields would be instrumental in identifying and incorporating these factors. 

This research is a substantial contribution to the exploration of quantum computing for mRNA codon optimization. The thoughtful recognition of potential areas for refinement underscores the depth of the research. Our resource-efficient quantum algorithm halves the required number of qubits in comparison to the existing approach \cite{fox2021mrna}.

\newpage
\appendix
\section*{Appendix: Hamiltonian for mRNA codon optimization}
In this appendix, we present detailed examples of constructing Hamiltonian for mRNA codon optimization using both one-hot encoding and dense encoding methods. The example includes a protein chain with GSK. 

\medskip
\noindent\textbf{\large{Terminology from the Cambridge dictionary}}
\medskip

\begin{itemize}[nolistsep,noitemsep]
    \item Nucleotides are a group of chemical compounds found in living cells in nucleic acids such as DNA and RNA.
    \item Polypeptides are a group of polymers made from a chain of amino acids.
\end{itemize}

\medskip
\noindent\textbf{\large{Synonymous Codon Encoding}}
\medskip

We leverage a dense encoding method to represent each synonymous codon for each amino acid in a protein chain. This method uses a string of qubits with each qubit assigned a value of `1' or `0' to define the mRNA sequence, while each synonymous codon has its own binary qubit to be represented. The number of required qubits for each amino acid is related to the number of synonymous codons available (equation \ref{dense}). Equation \ref{rna_qubits} represents the complete mRNA sequence for a protein chain, where $i$ represents the $i$th amino acid in a protein, and k represents the kth qubits. The choice of encoding method can significantly impact the optimization process and the resulting mRNA sequences.

\begin{equation}
    \label{rna_qubits}
    (q_0^1q_1^1 \cdot\cdot\cdot)(q_3^2q_4^2\cdot\cdot\cdot) \cdot\cdot\cdot (q_k^iq_{k+1}^i\cdot\cdot\cdot) \cdot\cdot\cdot (q_n^Nq_{n+1}^N\cdot\cdot\cdot)
\end{equation}

\medskip
\noindent\textbf{\large{Indicator}}
\medskip

To facilitate the construction of the Hamiltonian, an indicator is introduced for each synonymous codon. The indication streamlines the process of building Hamiltonian. For example, we define $Q_i(q)=f(q)$ with $3$ qubits return $1$ if the synonymous codon is chosen at position $i$ and return $0$ when the codon is not selected. The function is in equation \ref{dense_qubit}.
\begin{equation}
    \label{dense_qubit}
    \begin{split}
        Q_i(000) & = (1-q_k^i)(1-q_{k+1}^i)(1-q_{k+2}^i) \\
        Q_i(001) & = (1-q_k^i)(1-q_{k+1}^i)q_{k+2}^i \\
        Q_i(010) & = (1-q_k^i)q_{k+1}^i(1-q_{k+2}^i) \\
        Q_i(011) & = (1-q_k^i)q_{k+1}^iq_2^i \\
        Q_i(100) & = q_k^i(1-q_{k+1}^i)(1-q_2^i) \\
        Q_i(101) & = q_k^i(1-q_{k+1}^i)q_2^i \\
        Q_i(110) & = q_k^iq_{k+1}^i(1-q_2^i) \\
        Q_i(111) & = q_k^iq_{k+1}^iq_2^i \\ 
    \end{split}
\end{equation}

\medskip
\noindent\textbf{\large{Constructing the Hamiltonian}}
\medskip

The critical part of mRNA codon optimization involves the construction of the Hamiltonian, a fundamental element in the optimization process. The Hamiltonian takes into consideration several key factors that are essential for optimizing codons in a given protein sequence. These factors encompass codon usage bias, target GC concentration, codon context, and the presence of sequentially repeated nucleotides \cite{fox2021mrna}. Additionally, the Hamiltonian accounts for constraint items related to redundant encoding, as they cannot be translated to an actual mRNA sequence.

The primary objective is to identify the ground state with respect to the minimum value of the Hamiltonian. This minimum value is indicative of the optimal combination of codons for the given protein sequence. Once the minimum value is obtained, it can be decoded to derive the optimal mRNA sequence.

The Hamiltonian governing mRNA codon optimization can be represented by the following equation (equation \ref{Hamiltonian}):
\begin{equation}
    \label{Hamiltonian}
    H(q) = H_f(q) + H_{gc}(q) + H_r(q) + H_p(q)
\end{equation}
This Hamiltonian equation encapsulates the interplay of various factors that impact codon optimization. The task at hand is to systematically manipulate these elements to arrive at the minimal value, facilitating the generation of an optimal mRNA sequence that aligns with the desired codon usage, GC content, and nucleotide sequence characteristics.

This step in mRNA codon optimization is essential in the development of highly efficient mRNA sequences, catering to specific biological and functional requirements.

The Hamiltonian ($H$) used in mRNA codon optimization comprises several critical components, each contributing to the overall objective of determining the optimal mRNA sequence. These components are denoted as follows: $H_f$ for codon usage bias, $H_{gc}$ for target GC concentration, $H_r$ for sequentially repeated nucleotides, and $H_p$ for constraint items related to redundant encoding.

\medskip
\noindent\textbf{\large{Codon Usage Bias}}
\medskip

Codon usage bias plays a pivotal role in mRNA codon optimization. Different species exhibit preferences for specific codons when synthesizing proteins. The codons chosen in an mRNA sequence have a profound impact on both protein expression and mRNA folding. The optimization goal is to align the codon usage frequencies with those favored in a particular species. To achieve this, the objective function is designed to be inversely proportional to the frequency of codon usage. In essence, codons with higher usage frequencies should map to lower values of the Hamiltonian. The objective function for codon usage frequency ($H_f$) is expressed as follows (equation \ref{C_f}):

\begin{equation}
\label{C_f}
H_f = -c_f \sum ^{N-1} _{i=0} \log[c_i+\epsilon_f]Q_i
\end{equation}

Where $c_f$ is a tunable parameter, $c_i$ is the frequency of the codon at position $i$ in a specific host system, $\epsilon_f$ is a constant used to prevent infinite values, $Q_i$ is the dense encoding of the codon at position $i$, $N$ is the total number of codons.

\medskip
\noindent\textbf{\large{Target GC Concentration}}
\medskip

Target GC concentration indicates that the optimal mRNA sequence should closely match the desired GC content. This component of the Hamiltonian aims to minimize the difference between the actual GC content and the target GC content. The objective function for target GC concentration ($H_{gc}$) is expressed as follows (equation \ref{GC_concentration}):

\begin{equation}
\label{GC_concentration}
H_{gc} = c_{gc}(\sum_{i=0}^{N-1} s_{i}Q_{i} - N\rho_{gc})^2
\end{equation}

Where $C_{gc}$ is a tunable parameter.
$s_i$ represents the number of G and C nucleotides in the ith codon, $q_i$ is the binary encoding of the $i$th codon, $\rho_{gc}$ represents the target GC content.

\medskip
\noindent\textbf{\large{Sequentially Repeated Nucleotides}}
\medskip

This component aims to minimize the occurrence of sequentially repeated nucleotides within the mRNA sequence. The objective function for sequentially repeated nucleotides ($H_r$) is constructed as follows (equation \ref{repeated_H}):

\begin{equation}
\label{repeated_H}
H_r = C_r \sum_{i=0}^{N-2} r_{i}Q_{i}Q_{i+1}
\end{equation}

Where $C_{r}$ is a tunable parameter, $r_{i}$ represents the relative number of repeated nucleotides between sequential codons at positions $i$ and $i+1$, $q_i$ and $q_j$ represent the binary encoding of codons at positions $i$ and $j$, respectively.
Concrete examples of the calculation of $r_{i}$ are provided in the paper [Reference: fox2021mrna].

\medskip
\noindent\textbf{\large{Constraint Items for Redundant Encoding}}
\medskip

As previously mentioned, some binary encodings are considered redundant because they do not correspond to valid codons. To address this, the Hamiltonian includes constraint items to penalize redundant encodings, preventing them from being included in the ground state. The equation for the constraint items ($H_p$) is formulated as follows (equation \ref{C_p}):

\begin{equation}
\label{C_p}
H_{p} = c_{p}\sum_{i=0}^{N-1} Q_{i}
\end{equation}

These constraint items ($H_p$) serve to increase the value of unrelated encoding, effectively restricting the inclusion of redundant encodings in the optimal mRNA sequence.

The Hamiltonian, incorporating these components, is fundamental to the mRNA codon optimization process, guiding the selection of codons to achieve the desired biological and functional outcomes.

\newpage
\bibliographystyle{unsrt}
\bibliography{ref.bib}

\end{document}